\documentclass[twocolumn,showpacs,preprintnumbers,amssymb,aps,pra,]{revtex4}
\usepackage{amsmath}
\usepackage[dvips]{graphicx}
\usepackage[dvips]{color}
\usepackage{float}
\graphicspath{{figureseps/}} 

\begin{document}

\title{Creating atom-number states around tapered optical fibres by
  loading from an optical lattice}

\author{T.~Hennessy}
\email{thennessy@phys.ucc.ie}
\author{Th.~Busch}
\affiliation{ Department of Physics, University College Cork, Cork,
  Republic of Ireland}

\date{\today}

\begin{abstract}
  We describe theoretically a setup in which a tapered optical
  nanofibre is introduced into an optical lattice potential for cold
  atoms. Firstly, we consider the disturbance to the geometry of the 
  lattice potential due to scattering of the lattice lasers from the 
  dielectric fibre surface and show that the resulting distortion to 
  the lattice can be minimized by placing the fibre at an appropriate 
  position in the lattice. We then calculate the modifications of the 
  local potentials that are achievable by transmitting off-resonant light
  through the fibre. The availability of such a technique holds the
  potential to deterministically create and address small well-defined
  samples of atoms in the evanescent field of the tapered nanofibre.
\end{abstract}

\pacs{37.10.Jk,37.10.Vz,42.81.Wg}

\maketitle

\section{Introduction}
\label{sect:Introduction}
During the last two decades, advances in the cooling and trapping of
atoms and ions has assisted in the creation of clean and highly controllable
systems, in which fundamental quantum mechanical experiments can be
carried out with very low levels of noise. This has led to several
breakthrough successes in the quest for implementing ideas of
quantum information processing (QIP) \cite{Stolze:08},
high-precision atomic clocks \cite{Wilpers:04} and quantum
metrology \cite{Ye:08}.

For neutral atoms optical lattices have been important and hold a
great deal of promise in this area. The high degree of control one has
over the laser parameters has allowed for the execution of many
seminal experiments in these periodic systems. In particular, by
controlling the amplitude of the lasers one can adjust the trapping
depth, which can act as a switch between regimes in which the dynamics
are controlled either by tunnelling between different lattice sites or
by interactions between the atoms. This has led to the celebrated
observation of the superfluid-Mott insulator transition, in which a
state with one atom per lattice site can be created
\cite{Jaksch:98,Greiner:02}.

States which have a well-defined number of particles, so-called atomic
Fock states, are currently of large interest in physics. Their
sub-Poissonian number statistics is valuable for applications in
atom-metrology, quantum information processing and has merit for
investigating the foundation of quantum mechanics as well. Several
ground-breaking experiments have recently reported the creation of
such states \cite{Chuu:05,Esteve:08,Itah:10} and a significant amount
of theoretical work has been devoted to their characterization
\cite{Sokolovski:11,Pons:11}. Knowing the exact number of atoms a
priori in each run of the experiment is still a difficult task, and
techniques which can deterministically create a desired atom number
are under vigorous development.

Here we present a near-field optics approach to creating such definite
atom-number states and propose the use of the evanescent field of an
optical fibre as a tool for manipulating the optical lattice potential
locally (see Fig.~\ref{fig:Schematic}). While standard optical fibres 
have a diameter of several hundred $\mu$m, recent progress in tapering techniques allows for the
creation of fibres of subwavelength diameter \cite{Morrissey:09}
and even down to 50 nm\cite{Tong:03}. A significant amount of the
intensity in these fibres is carried in their evanescent field and can
therefore be used to create an optical potential for ultracold
atoms. In this work we will examine the effects of introducing a
sub-micrometer fibre into an optical lattice and demonstrate the
possibility of deterministically creating states of fixed particle
number using appropriately chosen fields inside the fibre.

The paper is organized as follows: in Sec.~\ref{sec:Potentials} we
will present a short review of the potential forces relating to
optical lattices and sub-micrometer diameter, single-mode silica
fibres. We then discuss the modifications of an optical lattice
potential due to effects from light scattering on a fibre in
Sec.~\ref{sect:Scattering} and examine several achievable potential
geometries resulting from the combination of the lattice potential and
the evanescent field potentials in Sec.~\ref{ModifiedLattice}. The
resulting atomic state is discussed in Sec.~\ref{sec:AtomicState} and
we finally conclude in Sec.~\ref{sec:Conclusions}.

\section{Potential Forces}
\label{sec:Potentials}

\subsection{Optical Lattices}
\label{subsect:optlatt}

To understand the influence the introduction of the fibre into an
optical lattice has, let us first briefly review the optical
potentials associated with optical lattices and nano-fibres. Optical
lattices exist by today in many laboratories and represent periodic
arrays of micro-traps generated by the dipole force of a standing wave
laser light field \cite{Jaksch:98,Greiner:02,Bloch:08}. A variety of
trapping geometries are achievable, with the most common being
rectangular \cite{Greiner:02} or triangular \cite{Becker:09}.

The simplest case of an optical lattice trapping potential is given by
a one-dimensional model, in which two counter-propagating laser beams
interfere. This results in a standing wave for the optical intensity
given by
\begin{equation}
  I(z)=I_0\sin^2(kz)\;,
  \label{eqn:lattice}
\end{equation}
where $k=2\pi/\lambda$ is the free space wave number of the laser
light, $I_0$ is the maximum intensity of the laser beam and the
periodicity is given by $\lambda/2$.  The spatially varying ac Stark
shift then forms a potential for the induced dipole moment, ${\bf p}$,
of the atom given by
\begin{equation}
  \label{eq:OpticalForce}
  U_\text{dip}=-\frac{1}{2}\langle{\bf p\cdot E}\rangle
              =-\frac{1}{2\epsilon_0c}\Re(\alpha)I
\end{equation}
where $\epsilon_0$ is the vacuum permittivity, $c$ the speed of light
and $\alpha(\omega_L)$ the optical polarizability, which depends on
the frequency of the laser field, ${\bf E}$. Its real part is given by
\cite{Jackson}
\begin{equation}
  \alpha(\omega)=2\pi\epsilon_0 c^3\sum_j\frac{g_j}{g_a}
   \frac{\gamma_{ja}(1-\frac{\omega^2}{\omega_{ja}^2})}
        {(\omega_{ja}^2-\omega^2)^2 + \gamma_{ja}^2 \omega^2},
        \label{eqn:atomicpolarizability}
\end{equation}
where the $g_j$ and $g_a$ are the statistical weights of the excited
and ground states, respectively, the $\omega_{ja}$ are the transition
frequencies and the $\gamma_{ja}$ are the emission transition
probabilities \cite{SANS}. Depending on the detuning of the laser
beam, the atoms can be forced to gather at the nodes or anti-nodes of
the laser intensity pattern by using light blue-detuned ${(\omega_L >
  \omega_0)}$ or red-detuned ${(\omega_L < \omega_0})$ with respect to
the chosen transition $\omega$, respectively.

\begin{figure}[tb]
  \begin{center}
    \includegraphics[width=1 \linewidth]{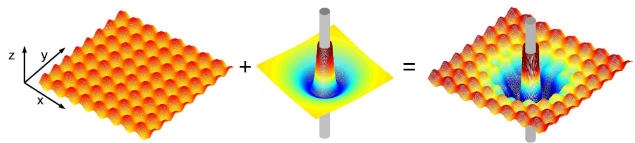}
  \end{center}
  \caption{(Color online) Schematic showing a typical optical potential geometry resulting from the
    combination of an optical lattice potential and a trapping
    potential around an optical nano-fibre.}
  \label{fig:Schematic}
\end{figure}

By introducing pairs of counterpropagating lasers in the remaining
directions of space, higher dimensional lattices can be created. The
interference terms between the laser fields in the different
directions can be eliminated by choosing perpendicular polarization
vectors of the two laser fields, which for a two-dimensional setup
results in an intensity pattern represented by the sums of purely
sinusoidal orthogonal fields (see Fig.~\ref{fig:Schematic})
\begin{eqnarray}
  I(x,y)=I_0\left[\sin^2(kx)+\sin^2(ky)\right]\;.
  \label{eqn:2Dlattice} 
\end{eqnarray}
Throughout this paper we will consider this type of two-dimensional
optical lattice, however a generalization to three dimensional,
layered lattices is straightforward. We will also assume that every
beam is independent and not created through retroreflection.

Optical Lattices typically have lattice constants in the range of
$400$ nm to $650$ nm and we will consider a lattice with a trapping
wavelength $\lambda/2=527$~nm. We choose the lattice to be loaded with
$^{133}$Cs atoms, which localize in the high field regions.

\subsection{Sub-wavelength diameter optical fibres}

Recent developments in tapered, dielectric fibre technology have made
it possible to produce fibres with radii, $a$, as low as a few hundred
nanometres \cite{Tong:03}. In such fibres the core has vanished and
they can be described by one large refractive index step between the
remaining cladding, $n_1(\omega)$, and the outside vacuum, $n_2$.  An
interesting consequence of the subwavelength nature of the diameter is
that the majority of the field will be guided in the evanescent field
on the fibre's surface. It therefore becomes accessible to atoms in
the fibre's vicinity and light blue-detuned with respect to the atoms
transition frequency will create a repulsive force preventing the
atoms from coming too close to the fibre which is at room-temperature.  At the
same time red-detuned light will result in an attractive force and a
combination of both fields was suggested by Le Kien {\it et al.}
\cite{LeKien:04} as a way of creating a trapping potential around the
fibre. This was experimentally observed in \cite{Vetsch:09}.

Let us briefly review the description of such a potential, following
closely \cite{LeKien:04}. We consider two frequencies,
$\omega_r$ and $\omega_b$, where the indices correspond to the red-
and blue-detuned fields, respectively. They are chosen such that the
single mode condition
\begin{equation}
  V_i\equiv k_i a \sqrt{n_1^2(\omega)-n_2^2 }< 2.405\;,
\end{equation}
is fulfilled \cite{Sellmeier} and both light fields are in the
fundamental mode HE$_{11}$.  The intensity distribution of the
evanescent fields depends on the polarization of the input fields and
here we choose circular polarization for both beams to achieve
angular symmetry \cite{Yariv:85,LeKien:04:2}. In cylindrical
co-ordinates $\{r,\phi,\theta\}$, the time-averaged intensity outside
the fibre is then given by 
\begin{equation} 
  |E_i|^2=\epsilon_i^2
          \left[K_0^2(q_i r)+w_iK_1^2(q_i r)+f_iK_2^2(q_i r)\right].
  \label{eqn:EFieldCircularOutside}
\end{equation}
Here the $K_n$ are the modified Bessel functions of the second kind
and $\epsilon _i$ is the strength of the electric field. The decay of
the fields from the surface of the fibre is characterized by $q_i$,
which is the reciprocal of the decay length $\Lambda_i$ and given by
\begin{equation}
  q_i=\sqrt{\beta_i^2-{n_2}^2 {k_i}^2},
\end{equation}
where $\beta$ is the longitudinal propagation constant of the mode
\cite{Yariv:85}.  Finally, the pre-factors are given by
\cite{LeKien:04:2}
\begin{align}
 w_i&=\frac{2{q_i}^2}{\beta^2_i(1-s_i)^2}\;,\\
 f_i&=\frac{(1+s_i)^2}{(1-s_i)^2}\;,
\end{align}
with $s$ defined as
\begin{equation}
  s_i=\frac{\left(\frac{1}{ q_i^2 a^2}+\frac{1}{h_i^2 a^2}\right)}
  {\left[\frac{J_1'(h_i a)}{h_i a J_1(h_i a)}
      +\frac{K_1'(q_i a)}{q_i a K _1(q_i a)}\right]}\;,
\end{equation}
and $h_i=(n_1^2 k_i^2-\beta_i^2)^{\frac{1}{2}}$.  The combined optical
potential around the fibre for a blue- and a red-detuned field is
therefore given by (see Fig.~\ref{fig:Schematic})
\begin{align}
  U(r)=&\frac{|\alpha_b|\varepsilon_b^2}{4}|
       \left[K_0^2(q_b r)+w_bK_1^2(q_b r)+f_bK_2^2(q_b r)\right] \nonumber \\
       &-\frac{|\alpha_r|\varepsilon_r^2}{4}|
       \left[K_0^2(q_r r)+w_rK_1^2(q_r r)+f_rK_2^2(q_r r)\right]\;,
      \label{eqn:twocolorfield}
\end{align}
where the factors in front of the mode-structure terms are directly
proportional to the powers of the individual light fields, $P_r$ and
$P_b$.

\subsection{Van der Waals interaction}
\label{subsect:vdW}

Finally we need to take into account the van der Waals attraction
between the atoms and the fibre. The classical van der Waals potential
felt by an atom near the surface of a dielectric fibre of infinite
length is given by \cite{Boustimi:02}
\begin{align}
  V(r)=\frac{\hbar}{4 \pi^3\epsilon_0} 
       \sum_{n=-\infty}^{\infty}& 
       \int_0^\infty dk[k^2K_n^{'}(kr)+(k^2 +\frac{n^2}{r^2})K_n^2(kr)]\nonumber \\ 
       &\times \int_0^{\infty} \!  d\xi \; \alpha(i \xi)G_n(i\xi)\;, 
       \label{eq:vdW}
\end{align}
where
\begin{equation}
  G_n(\omega)=\frac{[\epsilon(\omega)-\epsilon_0]I_n(ka)I_n'(ka)}
                 {\epsilon_0I_n(ka)K_n'(ka)-\epsilon(\omega) I_n'(ka)K_n(ka)}\;.
                   \label{eqn:GG}
\end{equation}
Here the $I_n(x)$ and $K_n(x)$ are the modified Bessel function of the
first and second kind, respectively. It should be noted that this
approximation neglects the resonant frequencies of silica. However, as
these are substantially different and weaker than those of Cs
atoms, this is justified \cite{LeKien:04}.

A detailed analysis of expression \eqref{eq:vdW} was carried out by
Le Kien {\it et al.} \cite{LeKien:04}, who found that for atoms close
to the surface the van der Waals potential tends to the same values as
that for a flat surface. The latter has the simple and well known form
\begin{align}
  \label{eq:vdWflat}
  V_\text{flat}=&-\frac{C_3}{(r-a)^3}\;,\\
  C_3=&\frac{\hbar}{16 \pi^2 \epsilon_0}
      \int_0^\infty \! d\xi\; \alpha(i \xi)  [ \frac {\epsilon(i \xi)-\epsilon_0}{\epsilon(i \xi)+\epsilon_0}]\;.    
\end{align}
The ground state Caesium atom has its dominant ($D_2$) line at 852 nm,
which gives a van der Waals constant of $C_3=2\pi\cdot1.56$kHz
$\mu$m$^3$ \cite{Lauraspaper} and this value will be used throughout
this paper. In the following we will also use the simplified
expression \eqref{eq:vdWflat} whenever justified while making sure the
full expression gives identical results.

\section{Scattering at the fibre}
\label{sect:Scattering}

When a fibre is introduced in a position perpendicular to the transverse plane of a
two-dimensional optical lattice, the four incident beams will be
scattered from the cylindrical surface and distort the regularity of
the lattice. To describe this we approximate the fibre by an infinite
cylinder of radius $a=150$ nm oriented orthogonally to the lattice
vectors and assume that the waves undergo a linear scattering process
\cite{Kerker:69}. In cylindrical co-ordinates the four incident waves
of the conventional lattice have the form of plane waves
\begin{equation}
  \textbf{E}_I(\theta_j)=\textbf{E}_0\exp\left[ikr\cos(\theta-\theta_j)\right],
\end{equation}
coming from the angles $\theta_j=0,\pi/2,\pi,3\pi/2$ and the total
lattice field is that given by
\begin{equation}
  E_I=\sqrt{\left(\textbf{E}_I(0)+\textbf{E}_I(\pi)\right)^2
           +\left(\textbf{E}_I\left(\frac{\pi}{2}\right)
                 +\textbf{E}_I\left(\frac{3\pi}{2}\right)\right)^2}.
\end{equation}
Assuming that the beams along the $\pi/2$-direction are polarized
parallel to the cylinder axis and the ones along the $\pi$-direction
orthogonal, the scattered field can be written as in \cite{Jones:73}, where the respective polarizations are contained in $E_0$.
  
\begin{align}
  E^{\parallel}_S=E_0\sum_{n=0}^\infty \Bigg[&
          i^na_nH_n^{(1)}(kr)
                \cos\left(n\left(\theta+\frac{\pi}{2}\right)\right)\nonumber\\
       &+ i^na_nH_n^{(1)}(kr)
                \cos\left(n\left(\theta+\frac{3\pi}{2}\right)\right)\Bigg]  \\
  E^{\perp}_S=\frac{E_0}{kr}\sum_{n=0}^\infty \Bigg[& 
          i^nb_nH_n^{(1)}(kr)\cos(n\theta) \nonumber \\
       &+ i^nb_nH_n^{(1)}(kr)\cos(n(\theta+\pi))\Bigg] 
\end{align}
where the $H_n^{(1)}$ are Hankel functions of the first kind and the
scattering coefficients are given by
\begin{align}
  a_n&=\frac{J_n(\alpha)J'_n(m\alpha)-mJ_n(m\alpha)J'_n(\alpha)}
            {H_n^{(2)}(\alpha)J'_n(m\alpha)-mJ_n(m\alpha)H_n^{(2)'}(\alpha)}\;,\\
  b_n&=\frac{m J_n(\alpha)J'_n(m\alpha)-J_n(m\alpha)J'_n(\alpha)}
            {mH_n^{(2)}(\alpha)J'_n(m\alpha)-J_n(m\alpha)H_n^{(2)'}(\alpha)}\;. 
\end{align}
From this the complete field follows as
\begin{equation}
     E_\text{tot}=
      \sqrt{\left(E^{\parallel}_I+E^{\parallel}_S\right)^2
           +\left(E^{\perp}_I+E^{\perp}_S\right)^2}\;. 
\end{equation}

\begin{figure}[tb]
  \includegraphics[width=\linewidth]{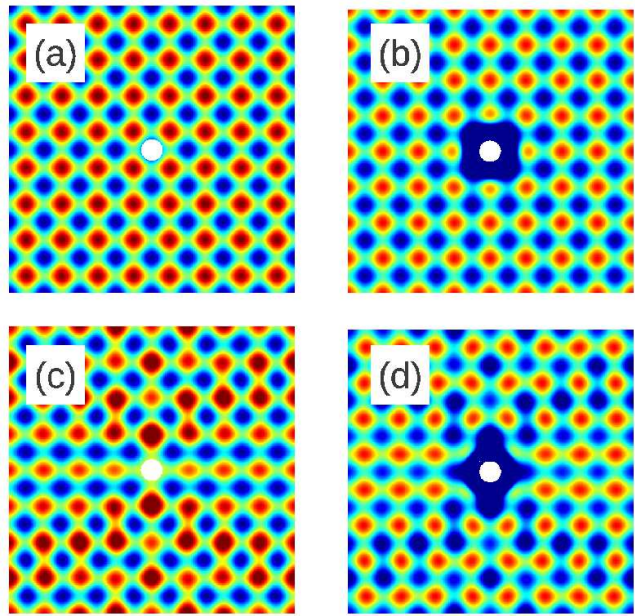}
  \caption{(Color online)(a) Lattice intensity including the
    scattering of the light on the fibre when the fibre is placed at
    an intensity minimum of the lattice. (b) van der Waals potential and optical lattice potential which includes the scattered lattice field. (c) and (d) show the same for a fibre
    placed at a lattice intensity maximum. Each plot spans an area of
    4.2 $ \mu$m $\times$ 4.2  $\mu$m and the lattice depth is 60 E$_R$. A colour scale has been used which varies from blue to red where the blue areas correspond to minima and red areas to maxima.}
   \label{fig:Scattering}
\end{figure}

Here we note that all of the square plots in this paper span an area of 4.2 $ \mu$m $\times$ 4.2 $ \mu$m. This 4.2 $ \mu$m corresponds to approximately four optical lattice wavelengths and thus an eight by eight grid of traps. The optical intensity in the vicinity of a fibre of radius $a=150$ nm
for a lattice constant of $\lambda/2=527$ nm is shown in
Figs.~\ref{fig:Scattering} (a) and (c).  One can see that if the fibre is located at a minimum of the optical intensity
(Fig.~\ref{fig:Scattering}(a)), the lattice structure is almost
unaffected. Positioning the fibre at an optical intensity maximum on
the other hand (Fig.~\ref{fig:Scattering}(c)) leads to noticeable
disturbances, which, nevertheless, leave the basic lattice structure
intact. Clearly larger fibres will lead to more scattering, however
the numbers chosen here are well in reach of experimental
possibilities.

In general the scattered radiation propagates as a cylindrical wave
and its intensity falls off as the inverse power of the radial
distance. Since the energy flow is only in the planes of constant $z$,
the scattered radiation corresponding to a particular incident ray
will be observed only in that plane which contains the incident ray
and no scattering into other layers of a three-dimensional lattice
occurs.

The overall potential seen by the atoms must include the van der Waals
potential and Figs.~\ref{fig:Scattering}(b) and (d) show that the
lattice sites most affected by the scattering are also strongly
affected by the van der Waals potential (Note that the Cs atoms we are
considering here are high field seekers).  It is clear that in a shallower
 lattice the effect of the van der Waals attraction will be
more severe on a larger range and we will show in the next section
that the introduction of a repulsive blue field can be a useful tool for
counteracting this effect.

\section{Adding fibre potentials}
\label{ModifiedLattice}
\subsection{Compensating the van der Waals potential}

\begin{figure}[tb]
  \includegraphics[width=8.75cm]{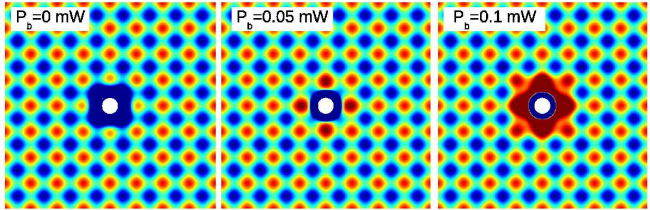}
  \caption{(Color online) Combined potentials (van der Waals,
    blue-detuned evanescent field and optical lattice) for a fibre
    placed at a minimum of the lattice intensity. The wavelength of
    the evanescent field is $\lambda_b=700$ nm and its power is
    increased through $P_b=$0, 0.05 and 0.10 mW from left to
    right. The lattice depth is chosen to be $=60E_\text{rec}$.}
   \label{fig:Fig3}
\end{figure}

\begin{figure}[tb]
  \includegraphics[width=8.75cm]{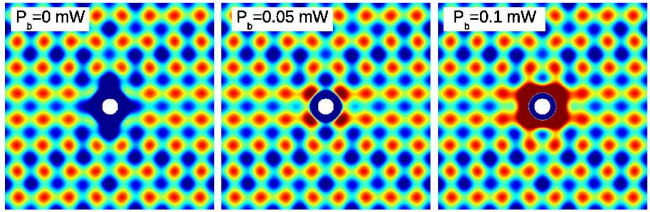}
  \caption{(Color online) Same as Fig.~\ref{fig:Fig3}
    with the fibre placed at a maximum of intensity of the lattice. Each plot spans an area of 4.2 $ \mu$m $\times$ 4.2  $\mu$m.}
   \label{fig:Fig4}
\end{figure}

In order to minimize the disturbance of the lattice due to the van der
Waals potential, we will study the possibility of compensating the
attractive potential with a repulsive one from a blue-detuned optical
field. The joint potential is simply given by adding the blue part of
eq.~\eqref{eqn:twocolorfield} to the van der Waals expression of
eq.~\eqref{eq:vdWflat}
\begin{align}
  U_{c}(r)=&\frac{|\alpha_b|\varepsilon_b^2}{4}\left[K_0^2(q_b r)+w_bK_1^2(q_b r)+f_bK_2^2(q_b r)\right] \nonumber \\
  &-\frac{C_3}{(r-a)^3}.
 \label{eqn:countervdwwithblue}
\end{align}
Since the modified Bessel functions have an exponentially decaying
form it is not possible to perfectly compensate the van der Waals
potential at all distances from the fibre. However, the discrepancy is
weaker at larger distances, which allows the reduction of the radius
in which the attractive potential is significant.  In
Figs.~\ref{fig:Fig3} and
\ref{fig:Fig4} we show the potential for the two
different positions of the fibre for different intensities of the blue
beam. One can clearly see that in both cases it is possible to achieve
a situation in which almost all lattice sites close to the fibre are
still intact (central panel). This is important for the construction
of a well-defined Mott insulator state around the fibre, which is a
prerequisite to loading the fibre potential with a well-defined
particle number. While the exact number of restored lattice sites also
depends on the lattice depth, the graphs show typical achievable
experimental values. For very small distances from the fibre surface,
the attractive van der Waals potential will always be stronger than the
compensating optical field and tunnelling will become an important loss
factor at longer times.

Let us also remark that with such a localized potential it is possible
to remove atoms from specific lattice sites by using the fibre as a
dark absorber \cite{Morrissey:09}. Used in conjunction with existing ideas involving
optical conveyor belts \cite{Schrader:01} this simple setup could be
an effective method of removing entire rows or patterns of atoms.

\subsection{Loading the fibre potential}

In the following we will consider the situation where an attractive,
red-detuned field is added to the fibre as well. This will allow for
the creation of a circular potential minimum around the fibre, deep
enough to trap ultracold atoms. Recent experiments have demonstrated
this by stochastically trapping atoms from a surrounding thermal or
condensed cloud \cite{Vetsch:09}. Since in our situation the
environment around the fibre is given by the well ordered optical
lattice, a controlled melting of the lattice by the evanescent field
will transfer a controllable number of atoms from the individual
lattice sites into the fibre potential. The resulting state is
therefore highly number squeezed and can be used in applications in
quantum information or metrology \cite{Esteve:08,delcampo}.

We study this process by assuming a realistic experimental situation
of a Mott insulator made from Cs atoms with a resonant transition
at $\lambda_0=852$ nm \cite{CesiumMott}.  For the two
light fields in the fibre we consider a blue-detuned field at a
wavelength of $\lambda_b=700$~nm and the red-detuned at
$\lambda_r=980$~nm. The detunings of the fibre fields from the
dominant line of the atom are then given by $\frac{\Delta_b}{2
  \pi}=-46$~THz and $\frac{\Delta_r}{2 \pi}=76$~THz and with a fibre
radius of $150$ nm, the evanescent decay lengths corresponding to the
blue and red fields are $\Lambda_b=0.36 ~\mu$m and $\Lambda_r=1.83
~\mu$m.  The two-dimensional optical lattice we consider has a depth 60$E_R$. 

\begin{figure}[tb]
\includegraphics[width=8.75cm]{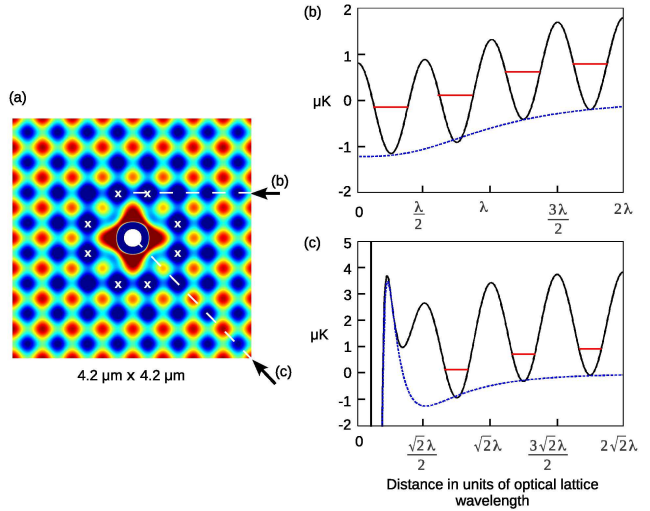}
\caption{(Color online) (a) The black solid line represents the
  resulting two-dimensional potential geometry if the red and blue
  detuned beams in the fibre are switched on for $P_b=0.12$ mW and
  $P_r=0.036$ mW. The lattice sites for which the energy is lowered
  such that the residing atoms will be trapped in the fibre potential
  after the optical lattice is switched off are marked with a
  cross. (b) and (c) cut through the potential at the lines indicated
  in (a). The relevant on-site energies are indicated as well and the
  dotted blue line indicates the potential after the optical lattice
  is switched off. The fact that the fibre potential does not always
  meet exactly the zero points of the overall potential before
  switch-off is due to phase shifting of the optical lattice after
  scattering on the fibre.}
\label{fig:Fig5}
\end{figure}

\begin{figure}[tb]
\includegraphics[width=8.75cm]{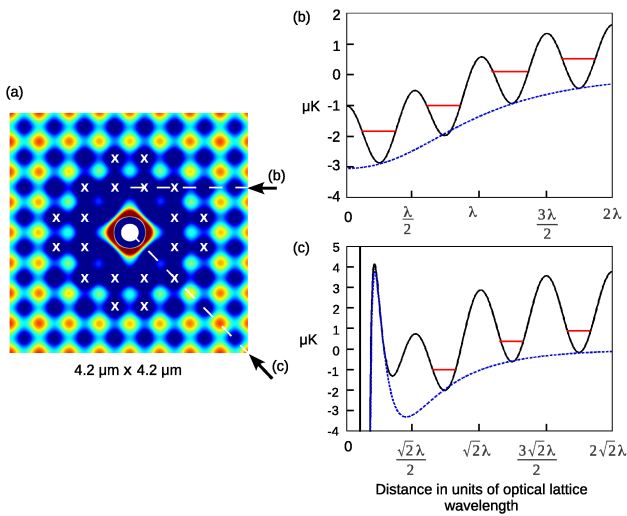}
\caption{(Color online) Same as Fig.~\ref{fig:Fig5}, but for higher
  powers of the evanescent fields, $P_b=0.17$ mW and $P_r=0.082$ mW. A
  larger number of atoms will be trapped after the optical lattice is switched off.}
\label{fig:Fig6}
\end{figure}

Two typical examples of resulting trapping geometries, when all
potentials are taken into account, are shown in Figs.~\ref{fig:Fig5}
and~\ref{fig:Fig6} for a fibre located at an intensity minimum. Since
we assume the fibre to be initially dark, the lattice sites which are
visible closest to the fibre in~Figs.~\ref{fig:Fig5}(c)
and~\ref{fig:Fig6}(c) are actually empty due to being within the
radius of the surface’s van der Waals potential. For all other sites
in the vicinity of the fibre one can clearly see that the addition of
the red and blue fields allows for the lowering of the on-site
energies. Therefore, as the asymptotic potential of the evanescent
field goes to zero and since there is no local maximum in the fibre
potential, a sudden switch off of the optical lattice will leave all
atoms with centre-of-mass energies $<0$ trapped in the fibre potential
alone. In the example shown in Fig.~\ref{fig:Fig5}, where $P_b=0.12$
mW and $P_r=0.036$ mW, one finds that this condition is fulfilled for
8 lattice sites. Increasing the evanescent fields to $P_b=0.17$ mW and
$P_r=0.082$ mW (see Fig.~\ref{fig:Fig6}) the radius of the evanescent
field increases and careful examination shows that 20 sites are
reached. Note that, due to the rectangular geometry of the considered
lattice, only certain atom numbers can be achieved and realistic
parameters limit this technique to samples of only a few tens. If the
switch-off process of the optical lattice is done on a time-scale
shorter than the typical atom tunnelling time in an optical lattice
(which is of the order of several milliseconds \cite{Greiner:02}), no
other atoms will have been able to join the well-defined sample.

The state created in this sudden-switch off procedure is a highly
excited, out-of-equilibrium state in the final potential and
subsequent cooling is necessary to prevent further atom loss due to
scattering and thermalisation. While a detailed calculation of these
effects go beyond the scope of this work, loss through
re-thermalisation can be minimised by applying a Feshbach resonance
while carrying out this process and subsequently adiabatically
lowering the fibre potential while switching the interaction back on
\cite{Alt:03}.
  
It is also worth pointing out that in both examples above the presence
of the repulsive blue field ensures the existence of a repulsive wall
between the fibre and the atoms, thereby preventing direct atom loss
through the room temperature object. However, the power in the blue
detuned field does not correspond to the same field strength that
optimally cancels the effects of the van der Waals potential, as
discussed in the previous chapter. It is rather necessary to
over-compensate the van der Waals potential and recreate the trapping
minimum using the red-detuned field. In the next section we will
discuss the nature of the atomic state created in the fibre potential.

{\color{red}{}} 

\section{Atomic State}
\label{sec:AtomicState}

Let us finally briefly characterise the atomic many-body state that
can be created by the procedure above and focus in particular on the
nature of the correlations in the sample. For this we first consider
the effective dimensionality of the ground state of the potential
around the fibre after the optical lattice is switched off. Since the
size of the radial ground state of the potential will be much smaller
than the curved, azimuthal one, we can assume an approximate
separation of the wavefunction in the two directions. This allows us
to describe the spectrum in the azimuthal direction by a free-space
periodic potential with the well known spectrum
$E_n^a=(n^2\hbar^2\pi^2)/(2mL^2)$, where $L=2\pi r_m$ is the
circumference of the potential at the position of its radial minimum,
$r_m$.  Since no analytical expression for the position of this
minimum is known, we find it numerically and estimate the energy
difference between the ground and the first excited state in the
azimuthal direction to be of the order of $\Delta
E_{10}^a=E_1^a-E_0^a\sim10^{-33}$ J, for both situations shown in
Fig.~\ref{fig:Fig5}

To find the spectrum in the radial direction, we numerically
diagonalize the radial part of the potential for a wide range of
parameters and find typical values for the separation of the ground
and first excited state to be of the order of $\Delta
E_{10}^r\sim10^{-30}$ J.  This significant difference in the stiffness
of the spectra in the two different directions (the $z$-direction can
be adjusted separately to be stiff) translates into an approximate
one-dimensional situation with an aspect ratio of $\sim 10^3$. 

Having established the effective dimensionality of the potential, the
many body state of a one-dimensional Bose gas can now be characterized
using the Lieb-Liniger parameter $\gamma=mg_{1D}/\hbar^2 n_{1D}$
\cite{Lieb:63}. Here $g_{1D}$ is the one-dimensional coupling constant
given by $g_{1D}=\frac{4 \hbar^2 a_{3D}}{m
  a_\perp}(a_\perp-Ca_{3D})^{-1}$ with $C\approx 1.4603$ and $n_{1D}$
is the linear density of the atoms \cite{Olshanii:98}. For values of
$\gamma\gg 1$ the atomic many body state would be in the strongly
correlated Tonks-Girardeau regime, whereas for $\gamma<1$ the gas can
be treated as weakly correlated. For both cases shown in
Fig.~\ref{fig:Fig5} the radial ground state size is of the order of
$a_\perp\sim 0.2\;\mu$m (we assume the same is achieved in the
z-direction) and the position of the radial minimum is at
$r_m\sim6.6\;\mu$m. This leads to values of $\gamma=  0.560$ for the 8
atom case and $\gamma= 0.346$ for the 20 atom case, putting both
states firmly in the weakly correlated regime.

\section{Conclusion}
\label{sec:Conclusions}

In this work we have suggested that the combination of optical
lattices and tapered optical nano-fibres can be used to create small
atomic samples which allow control over the final atom number. While
introducing the fibre into the optical lattice inevitably leads to a
disturbance of the lattice in the vicinity of the fibre due to
scattering of the lattice beams, we have shown that this can be
minimized and, due to the small fibre diameter, usually leaves the
overall lattice structure intact. The attractive van der Waals
potential close to the surface of the fibre can be compensated by
using a blue-detuned evanescent field around the fibre, which allows
a reduction of the range of the fibre's influence to the size of a
single lattice site for typical experimental parameters. Adding a
second, red-detuned light field to the fibre then allows local melting
of the optical lattice and can be used to create a small sample with a
well-defined atom number. Finally, we have shown that these samples
are in the superfluid regime and therefore good candidates for
investigating the physics of persistent currents, or, using more
varied optical potentials around the fibre, the physics of superfluid
squids.

\begin{acknowledgements}
  This project was supported by Science Foundation Ireland under
  project numbers 05/IN/I852 and 10/IN.1/I2979. We would like to thank S\'ile Nic
  Chormaic, Laura Russell, Mary Frawley, David Rea and Vladimir Minogin for valuable
  discussions.
\end{acknowledgements}


\begin{thebibliography}{99}
% Quantum Computing: A Short Course from Theory to Experiment
\bibitem{Stolze:08} J.~Stolze and D.~Suter, {\sl Quantum Computing: A
Short Course from Theory to Experiment}, Wiley VCH, (2008).

% Optical Clock with Ultracold Neutral Atoms
\bibitem{Wilpers:04} G.~Wilpers, T.~Binnewies, C.~Degenhardt,
U.~Sterr, J.~Helmcke, and F.~Riehle, Phys.~Rev.~Lett.~{\bf 89}, 230801 (2002).

% Quantum State Engineering and Precision Metrology Using
% State-Insensitive Light Traps
\bibitem{Ye:08} J.~Ye, H.J.~Kimble, and H.~Katori, Science {\bf 320}, 1734 (2008).

%Cold Bosonic Atoms in Optical Lattices
\bibitem{Jaksch:98} D.~Jaksch, C.~Bruder, J.I.~Cirac, C.W.~Gardiner,
and P.~Zoller, Phys.~Rev.~Lett.~{\bf 81}, 3108 (1998).
 
% Quantum phase transition from a superfluid to a Mott insulator in a gas of ultracold atoms
\bibitem{Greiner:02} M.~Greiner, O.~Mandel, T.~Esslinger, T.W.~H\"ansch, and I.~Bloch, Nature {\bf 415} 39 (2002).

%Slow Mass Transport and Statistical Evolution of an Atomic Gas across the SuperfluidMott-Insulator Transition
\bibitem{CesiumMott} Chen-Lung Hung, Xibo Zhang, Nathan Gemelke, and Cheng Chin, Phys. Rev. Lett. \textbf{104}, 160403 (2010)

% Direct Observation of Sub-Poissonian Number Statistics in a
% Degenerate Bose Gas
\bibitem{Chuu:05} C.-S.~Chuu, F.~Schreck, T.P.~Meyrath, J.L.~Hanssen,
G.N.~Price, and M.G.~Raizen, Phys.~Rev.~Lett.~{\bf 95}, 260403 (2005).

% Squeezing and entanglement in a Bose-Einstein condensate
\bibitem{Esteve:08} J.~Esteve, C.~Gross, A.~Weller, S.~Giovanazzi, and
M.K.~Oberthaler, Nature {\bf 455}, 1216 (2008).

% Direct Observation of a Sub-Poissonian Number Distribution of
% Atoms in an Optical Lattice
\bibitem{Itah:10} A.~Itah, H.~Veksler, O.~Lahav, A.~Blumkin,
  C.~Moreno, C.~Gordon, and J.~Steinhauer, Phys.~Rev.~Lett.~{\bf 104},
  113001 (2010).

% Atomic Fock states by gradual trap reduction: From sudden to
% adiabatic limits
\bibitem{Sokolovski:11} D.~Sokolovski, M.~Pons, A.~del Campo, and
  J.G.~Muga, Phys.~Rev.~A {\bf 83}, 013402 (2011).

% Fidelity of fermionix atom-number states subjected to tunneling
% decay
\bibitem{Pons:11} M.~Pons, D.~Sokolovski and A.~del Campo,
  arXiv:1111.1346 (2011).

% Tapered optical fibers as tools for probing magneto-optical trap
% characteristics
\bibitem{Morrissey:09} M.J.~Morrissey, K.~Deasy, Y.~Wu, S.~Chakrabarti
and S.~Nic Chormaic, Rev.~Sci.~Instrum.~{\bf 80}, 053102 (2009).

 % Subwavelength-diameter silica wires for low-loss optical wave
 % guiding
\bibitem{Tong:03} L.~Tong, R.R.~Gattass, J.B.~Ashcom, S.~He, J.~Lou,
 M.~Shen, I.~Maxwell, and E.~Mazur, Nature {\bf 426}, 816 (2003).

% Many-body physics with ultracold gases.
\bibitem{Bloch:08} I.~Bloch, J.~Dalibard, and W.~Zwerger,
Rev.~Mod.~Phys.~{\bf 80}, 885 (2008).

% Ultracold quantum gases in triangular optical lattices
\bibitem{Becker:09} C.~Becker, P.~Soltan-Panahi, J.~Kronj\"ager,
S.~D\"orscher, K.~Bongs, and K.~Sengstock, arXiv:0912.3646 (2009).

%Quantum computation in optical lattices via global laser addressing
%\bibitem{superlattice} A. Kay and J. K. Pachos, New J. Phys. \textbf{6} 126 (2004);


\bibitem{Jackson} See, for example, J.D. Jackson, \textit{Classical
    Electrodynamics}, 3rd ed. (John Wiley \& Sons, New York, 1998).

  % Handbook of Basic Atomic Spectroscopic Data
\bibitem{SANS} J.E. Sansonetti, W.C. Martin, and S.L. Young (2005),
  Handbook of Basic Atomic Spectroscopic Data (version
  1.1.2). [Online] Available: http://physics.nist.gov/Handbook [2010,
  03 01]. National Institute of Standards and Technology,
  Gaithersburg, MD.

% Optical interface created by laser-cooled atoms trapped in the
% evanescent field surrounding an optical nanofiber
\bibitem{Vetsch:09} E.~Vetsch, D.~Reitz, G.~Sagu\'e, R.~Schmidt,
 S.T.~Dawkins, A.~Rauschenbeutel, Phys. Rev. Lett. \textbf{104}, 203603 (2010) 
%arxiv:0912.1179 (2009).



  % Atom trap and waveguide using a two-color evanescent light field
  % around a subwavelength-diameter optical fibre.
\bibitem{LeKien:04}F.~Le Kien, V.I.~Balykin, and K.~Hakuta,
  Phys.~Rev.~A {\textbf 70}, 063403 (2004).



\bibitem{Sellmeier} The refractive index $n_1$ of fused silica
  $(SiO_2)$ can be calculated using a Sellmeier-type dispersion formula
  \cite{LeKien:04}, taking the refractive index of the vacuum $n_2 =1$
\begin{align}
  n_1-1=&\frac{0.696166\lambda^2}{\lambda^2-(0.068404)^2}
        +\frac{0.407942\lambda^2}{\lambda^2-(0.116241)^2}\nonumber\\
       &+\frac{0.897479\lambda^2}{\lambda^2-(9.896161)^2}\nonumber
\end{align}
where $\lambda$ is in units of $\mu$m.

  % Optical Electronics.
\bibitem{Yariv:85} A.~Yariv, {\em Optical Electronics} ,CBS College,
  New York (1985).

  % Field intensity distributions and polarization orientations in a
  % vaccum-clad subwavelength-diameter optical fibre.
\bibitem{LeKien:04:2} F.~Le Kien, J.Q.~Liang, K.~Hakuta, and
  V.I.~Balykin, Opt.~Comm.~{\bf 242}, 445 (2004).

  % van der Waals interaction between an atom and a metallic nanowire
\bibitem{Boustimi:02} M.~Boustimi, J.~Baudon, P.~Candori, and
  J.~Robert, Phys.~Rev.~B {\bf 65}, 155402 (2002).

  % Molecules interacting with a metallic nanowire
%\bibitem{Boustimi:03} M.~Boustimi, J.~Baudon, and J.~Robert,
%  Phys.~Rev.~B {\bf 67}, 045407 (2003).

%Spectral Distribution of atomic flouresence coupled into an optical nano-fibre.
\bibitem{Lauraspaper} L.Russell, D.A.Gleeson, V.G.Minogin, and S.Nic Chormaic, J.Phys.B: At. Mol. Opt. Phys. \textbf{42}, (2009), 185006 (9pp).

% The Scattering of Light: and other electromagnetic radiation.
\bibitem{Kerker:69} M.~Kerker, {\em The Scattering of Light: and other
electromagnetic radiation},Academic. Press, N.Y. and London (1969).

% Light scattering by a cylinder situated in an intereference
% pattern with relevance to fringe anemometry and particle sizing.
\bibitem{Jones:73} A.R.~Jones, J.~Phys.~D:~Appl.~Phys.~{\bf 6}, 417
  (1973).


  % An optical conveyor belt for single neutral atoms
\bibitem{Schrader:01} D.~Schrader, S.~Kuhr, W.~Alt, M.~M\"uller,
  V.~Gomer and D.~Meschede, Appl.~Phys.~B {\bf 73}, 819 (2001).

\bibitem{Alt:03} W.~Alt, D.~Schrader, S.~Kuhr, M.~Muller, V.~Gomer, and D.~Meschede. Phys.~Rev.~A, {\bf{67}}, 033403 (2003).
%“Single atoms in a standing-wave dipole trap,” 



%Fidelity of fermionic atom-number states subjected to tunneling decay
\bibitem{delcampo} M. Pons, D. Sokolovski and A. del Campo, New J. Phys. \textbf{12} (2010) 065025

\bibitem{Alt:03} W.~Alt, D.~Schrader, S.~Kuhr, M.~Muller, V.~Gomer, and D.~Meschede. Phys.~Rev.~A, {\bf{67}}, 033403 (2003).
%“Single atoms in a standing-wave dipole trap,” 


  % Exact Analysis of an Interacting Bose Gas
\bibitem{Lieb:63} E.~Lieb and W.~Liniger, Phys.~Rev.~{\bf 130}, 1605
  (1963); E.~Lieb, Phys.~Rev.~{\bf 130}, 1616 (1963).

  % Atomic Scattering in the Presence of an External Confinement and a
  % Gas of Impenetrable Bosons
\bibitem{Olshanii:98} M.~Olshanii, Phys.~Rev.~Lett.~{\bf 81}, 938
  (1998).


\end{thebibliography}
\end{document}